\documentstyle[12pt]{article}
\newcommand{\newc}{\newcommand}
\newc{\beq}    {\begin{equation}}
\newc{\eeq}    {\end{equation}}
\newc{\beqa}    {\begin{eqnarray}}
\newc{\eeqa}    {\end{eqnarray}}
\newc{\bibi}    {\bibitem}

\begin{document}

\begin{titlepage}
\title{
 Open Inflation With  Scalar-tensor Gravity }

\author{ Jae-weon Lee,\\
 Seoktae  Koh \\
 and\\
 Chul H. Lee,\\
      \it Department of Physics,
      \\ \it          Hanyang University,
                Seoul, 133-791, Korea
        \\
\\
}

\date{}
\setcounter{page}{1}
\maketitle
\vspace{-15cm}
\hfill 
\vspace{13cm}

The open inflation model recently proposed
 by Hawking and Turok
is investigated in  scalar-tensor gravity context.
If the dilaton-like field has no potential,
 the instanton of our model is singular but has a finite action.
 The Gibbons-Hawking surface term vanishes
 and hence, can not be used to make $\Omega_0$   nonzero.
To obtain a successful open inflation one should introduce
 other matter fields or
a potential for the dilaton-like fields. \\
\vspace{1cm}
PACS numbers: 98.80.Cq
\maketitle

\end{titlepage}

\newpage

 Recently, Hawking and Turok \cite{HT,HT2}  proposed that
an open inflation can be obtained by the
singular instanton describing
quantum creation of a homogeneous open
universe with `no-boundary' proposal\cite{HH}.
Their model does not require  fine tuning
of parameters  to obtain the bubble formation and, at the same time,
the slow-roll inflation required in earlier models\cite{BGT}.
However, the HT(Hawking-Turok) instanton solution is singular
and the physical nature of this singularity is still controversial
\cite{singular,linde,bousso}.

In this paper, we will extend their works to the case with
  scalar-tensor gravity.
Since the quantum creation scenario of the universe
is adequate at the Planck scale,
it is natural to consider the extended gravity sector
which is common to  the unified theories
such as supergravity, superstring and Kaluza-Klein theory\cite{sugra}.
For example, recently, effective low-energy four dimensional
 Lagrangians have been obtained from spherical compactifications of
 string/M-theory\cite{m}.

In the no-boundary quantum cosmology, the probability of creation
of an universe is given by $P\propto Exp(-2S_E)$, where $S_E$ is
the Euclidean four action of the instanton. In the simplest version
of HT's model the most probable universe is that with the present
density parameter $\Omega_0=0$. Hence, it was required to introduce
the anthropic principle or other matter fields to obtain the
nonzero $\Omega_0$. The Gibbons-Hawking surface term  also
contribute to $S_E$ and play a role in making $\Omega_0$ nonzero.

We here investigate how all these facts change when we consider
the instanton with  scalar-tensor gravity.

Let us find the instanton solution of our model.
The  general Euclidean $O(4)$ symmetric metric is
\beq
ds^2=d\tau^2+b^2(\tau)d\Omega_3^2,
\label{metric}
\eeq
where $d\Omega^2_3$ is the metric describing $S^3$
and $b$ is the cosmic scale factor.
We consider the Euclidean four action which is given by\cite{acceta}
\beq
S_E=-\int \sqrt{g} d^4x[  \xi \phi^2 R
-\frac{1}{2}\partial_\mu\phi \partial^\mu \phi -U(\phi)
-\frac{1}{2}\partial_\mu\sigma \partial^\mu \sigma -V(\sigma)],
\label{action}
\eeq
where $\phi$ is a dilaton or Brans-Dicke like field and $\sigma$ is
an inflaton field.
In this frame $\phi$ couples to gravity non-minimally.
For simplicity, we will set $U(\phi)=0$.
Then our gravity sector
corresponds to the ordinary Brans-Dicke theory after a
conformal rescaling. In this case Brans-Dicke
parameter $\omega$, which is constrained to be greater than $500$
by experiment, is equal to $1/ 8\xi $. Hence, $\xi\ll1$.

From $S_E$ we obtain the  equations of motion for $b$ and the scalar fields
\beqa
\frac{\dot{b}^2}{b^2}&=&-\frac{2\dot{b}\dot{\phi}}{b\phi}+\frac{1}{b^2}+
\frac{1}{6\xi\phi^2}[\frac{\dot{\phi}^2}{2}+\frac{\dot{\sigma}^2}{2}
-V(\sigma)],
\label{b}
\\
\ddot{\phi}&=&-\frac{3\dot{b}\dot{\phi}}{b}-\frac{\dot{\phi}^2}{\phi}-
\frac{1}{1+12\xi}[\frac{\dot{\sigma}^2}{\phi}+\frac{4}{\phi}
V(\sigma)],
\label{phi}
\\
\ddot{\sigma}&=&-\frac{3\dot{b}\dot{\sigma}}{b}+V'(\sigma),
\label{sigma}
\eeqa
where  the dots denote the derivative with respect to
$\tau$ and $V'(\sigma) \equiv dV/d\sigma$.
The instanton is a solution of the above equations
with boundary conditions:\\
$b=0,\dot{b}=1,$ and
$\dot{\phi}=0=\dot{\sigma}$ at $\tau=0$
and $b=0$ again at some later $\tau=\tau_f$.
 Fig.1. shows  numerical solutions of the equations
 with $V(\sigma)=m^2 \sigma^2$, where the inflaton mass $m= 10^{13}~GeV$.

It is nontrivial to show the full behaviour of
the system. Hence, we will restrict ourselves to the
monotonically increasing and then decreasing
 solutions where $b$ increases maximally ($\dot{b}=0$) at $\tau=\tau_{m}$
  and $b=0$ again at $\tau=\tau_f$ as shown  in Fig.1.
Except for the first term the right hand side of Eq.(\ref{phi})
 always  makes negative
contribution to $\ddot{\phi}$.
However, for $\tau<\tau_{m}$  ($\dot{b}>0$)
 this term can give only  damping force to $\phi$,
 and $\dot{\phi}\le 0$.
For $\tau\ge \tau_{m}$, since  $\dot{b}\le 0$ and
by continuity, $\dot{\phi}$ remains negative.
 Hence $\phi$ is a monotonically decreasing
function of $\tau$ and becomes zero at the singularity($\tau=\tau_f$).

The equation for $\sigma$ is the same as
that with Einstein gravity.
So $\phi$ can have an effect on
$\sigma$ only through $b(\tau)$. This implies that
it is nontrivial to avoid the singular behaviour of the HT instanton
near $\tau=\tau_f$, even with  scalar-tensor gravity.
Near the singular region and for the sufficiently flat potential
 we can ignore the $V(\sigma)$ and $V'(\sigma)$
dependent terms, and $\dot{\sigma} $ goes like $b^{-3}$.
Since $\dot{b}^2\ge 0$ and $\phi$ approaches  zero
 near the singularity, we expect
\beq
\frac{b(\dot{\phi}^2+\dot{\sigma}^2)}{12\xi\phi^2}
 \gg \frac{2\dot{b}\dot{\phi}}{ \phi},
\label{approx}
\eeq
 if there is no miraculous cancellation
between the terms in Eq.(\ref{b}).

On the other hand, from the field equations one can obtain the following equation
for $b$:
\beq
\ddot{b}=2\frac{\dot{b}\dot{\phi}}{\phi}-\frac{b}{6\xi\phi^2}
[ \dot{\phi}^2+\frac{1+6\xi}{1+12\xi}{\dot{\sigma}}^2
+\frac{1-12\xi}{1+12\xi} V(\sigma)].
\eeq
In the case where $\xi\ll 1$, using Eq.(\ref{b}) and  Eq.(\ref{approx})  the
above equation  reduces to
\beq
\ddot{b}\simeq-\frac{2 \dot{b}^2}{b},
\eeq
which has a solution $ b(\tau)\propto (\tau_f-\tau)^{\frac{1}{3}}$
like in the HT instanton.
With Eq.(\ref{sigma}) this implies $\sigma(\tau)\propto
ln(\tau_f-\tau)$.

With the metric in Eq.(\ref{metric}) and the $O(4)$ symmetric fields,
 $S_E$ is given by
\beq
S_E=\pi^2\int d\tau\{b^3[\frac{\dot{\phi}^2}{2}+\frac{{\dot{\sigma}}^2}{2}
+V(\sigma)]+6\xi\phi^2(\ddot{b}b^2+{\dot b }^2b-b)\},
\label{const}
\eeq
where we have inserted $R=-6b^{-2}(b\ddot{b}+{\dot{b}}^2-1)$ into
the equation
and the integration was taken  over the half of $S^3$.
Using Eq.(\ref{b}) in Eq.(\ref{const})
 and integrating by parts, we obtain
\beq
S_E=\pi^2\int^{\tau_f}_{0} d\tau(b^3V(\sigma) -6\xi\phi^2b)+
2\pi^2\xi(b^3\dot{)}\phi^2|_{\tau=\tau_f},
\label{SE2}
\eeq
where the last term  should be canceled by
  the Gibbons-Hawking surface term\cite{HG}.
In the approximation  of constant fields
 ($\phi=\phi_0$,~$\sigma=\sigma_0$) and
the $O(5)$ symmetry ($b(\tau)\simeq H_0^{-1} Sin(H_0\tau)$),
the integration yields the usual factor,\\
$-12\pi^2 M_P(\phi_0)^4/V(\sigma_0)$. Here
$M_p(\phi_0)\equiv (2\xi\phi_0^2)^{\frac{1}{2}}$ is the reduced Planck mass
corresponding to $\phi_0$ and $H_0$ is the corresponding Hubble parameter.
The quantities with the subscript $0$ are the values at
$\tau=0$.
If we consider only the  first term in $S_E$,
it has a minima at
  $\sigma_0=0$ and the most probable universe is
that with $\Omega_0=0$.

Near the singularity,
since  $ b(\tau)\propto (\tau_f-\tau)^{\frac{1}{3}}$,
the integrand in Eq.(\ref{SE2}) does not diverge
 and gives finite contribution to the action.
Since $\phi$ goes to zero as $\tau$
approaches  $\tau_f$,
the surface term vanishes and
 can not be used to shift
 the minima of $S_E$ and to get nonzero $\Omega_0$.
Therefore, the Euclidean action of the instanton is finite.

What is the role of the four-form field
 $F_{\mu\nu\rho\lambda}$ in our model?
Since the role of the four-form fields
is to give a constant contribution
to the cosmological constant and the energy momentum tensor,
the generic behavior of the other fields does not change.

Adding
\beq
S_F=\int d^4x \frac{\sqrt{g}} {48}
F_{\mu\nu\rho\lambda}F^{\mu\nu\rho\lambda}
\eeq
into $S_E$, one can get
\beq
S_E\simeq -12\pi^2 \frac{M_p^4(\phi_0)}{V(\sigma_0)^2}
 [V(\sigma_0)-\frac{F^2}{48}],
\label{sol}
\eeq
where $F^2=F_{\mu\nu\rho\lambda}F^{\mu\nu\rho\lambda}$.

However, one must  add a total divergence term
to obtain a stationary action under
variations where the four-form is fixed on the boundary.
This term can cancel the $F^2$ term\cite{HT2,linde,4form,duncan}.
So, including the four-form field one can not obtain the
desired $\Omega_0$ value. But, the field can play a role
in the cosmological constant problem.

The vanishing of $\phi$ at the singularity
 means that the effective gravitational
constant diverges there and we could not ignore non-perturbative
effects or higher order terms,
 if our theory is a low energy effective one of some
fundamental theory.
Hence, it might be necessary to consider the effects of
 nonzero $U(\phi)$, higher order correction terms,
 and other matter fields\cite{HT2,garriga,garriga2,quad}.

In summary, in the context of the Brans-Dicke like
gravity, we found the finite action instanton which is singular
and has a vanishing Gibbons-Hawking surface term.
Similar to  HT's model, our model  requires additional
matter fields or a dilaton potential to obtain a successful open
inflation.

\vskip 1cm
 After completion of this paper, we became aware of
 a related work in the Einstein frame\cite{string}.
 Authors  are thankful to Dr. Minho Lee and Dr. Hyungchan Kim
 for useful discussions.
We are also thankful to Prof. A. Linde
who pointed out us that
the four-form field can not play a role in
making $\Omega_0$ nonzero in our model.
This work was supported in part by KOSEF.

\section*{ Figure Caption }
Fig.1 Numerical solutions of the field equations
with $\xi=5\times 10^{-5}$. The scale of $b(\tau)$ is reduced by $50$.
The mass unit is the reduced Planck mass at present.

\newpage


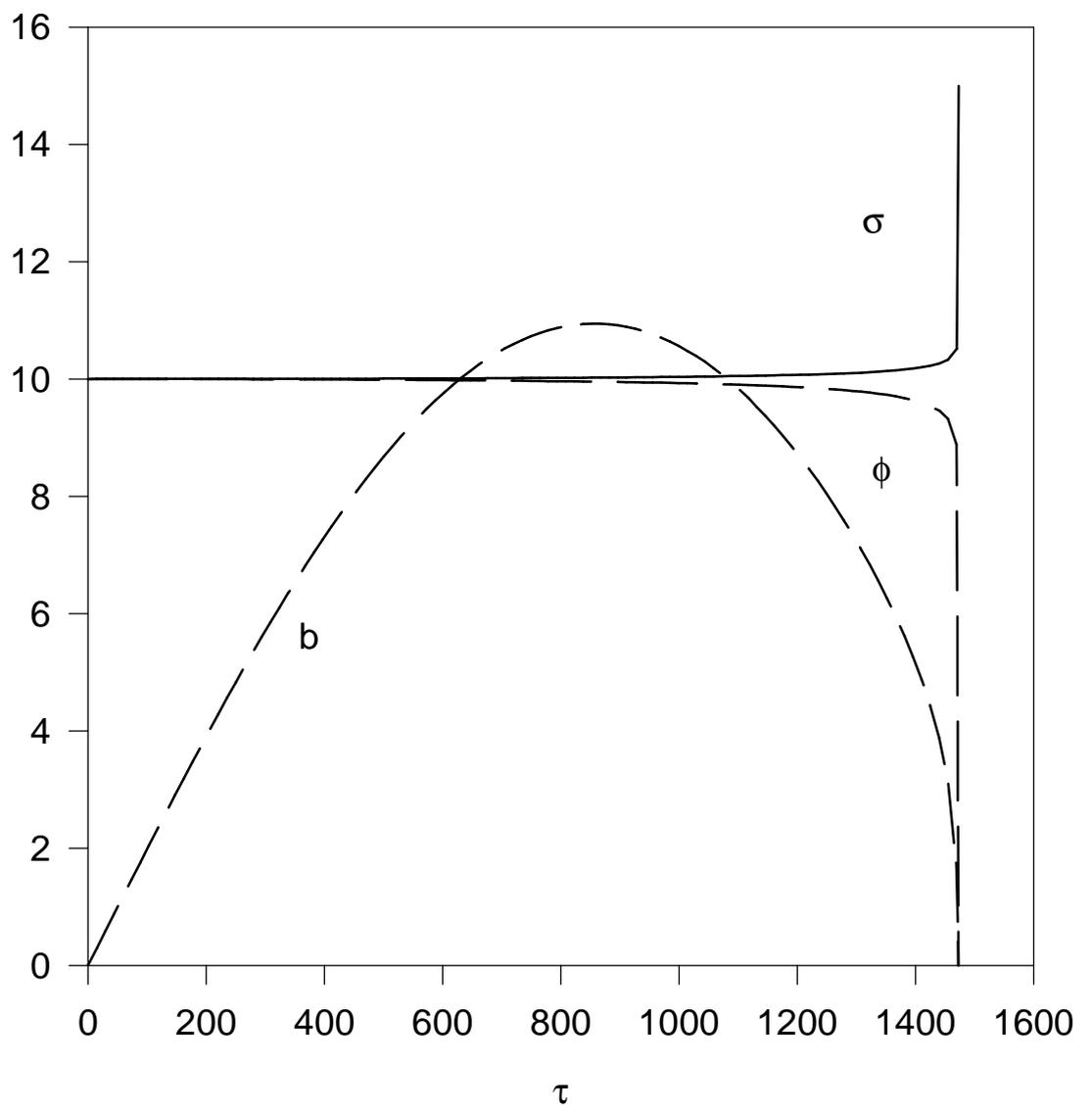
\end{document}